\documentclass[review]{elsarticle}

\usepackage{lineno,hyperref}
\modulolinenumbers[5]
\usepackage{graphicx,color}
\usepackage{wrapfig}
\usepackage{dcolumn}
\usepackage{gensymb}

\journal{arXiv}

\begin{document}

\begin{frontmatter}

\title{Detecting magneto-optical interactions in nanostructures}

\author{Luana Hildever$^{1}$, Thiago Ferro$^{1}$, Adrielson Dias$^{2}$, André José$^{2}$, Francisco Estrada$^{3}$, and José Holanda$^{1, 2, }$\corref{mycorrespondingauthor}}
\cortext[mycorrespondingauthor]{Corresponding author: joseholanda.silvajunior@ufrpe.br}
\address{$^{1}$Programa de Pós-Graduação em Engenharia Física, Universidade Federal Rural de Pernambuco, 54518-430, Cabo de Santo Agostinho, Pernambuco, Brazil.}
\address{$^{2}$Unidade Acadêmica do Cabo de Santo Agostinho, Universidade Federal Rural de Pernambuco, 54518-430, Cabo de Santo Agostinho, Pernambuco, Brazil.}
\address{$^{3}$Facultad de Biología, Universidad Michoacana de San Nicolas de Hidalgo, Av. F. J. Mujica s/n Cd. Universitaria, Morelia, Michoacán, México.}

\begin{abstract}
	Effects due to magneto-optical interactions are responsible for most of the phenomena discovered in optoelectronics and spintronics. Magneto-optical interactions can generate elementary excitations of the order of light-magnetic matter, which can flow under certain conditions. Here, we observe the intensities of magneto-optical interactions in hexagonal arrays of magnetic nanowires using experimental measurements and simulations. Nanowires of three materials (cobalt-Co, iron-Fe, and nickel-Ni) were electrodeposited on alumina membranes by the AC electrodeposition method. Our results reveal that the magneto-optical behavior can produce, under certain conditions, a kind of avalanche of magneto-optical interactions, which is dynamic. Such an observation shows the possibility of generating a magneto-optical current (spin-opto current). 
\end{abstract}

\begin{keyword}
	\texttt{}spintronics\sep optoeletronics\sep magneto-optical\sep interactions
\end{keyword}
\end{frontmatter}
\vspace{1.5cm}

\section{Introduction}

Magneto-optical effects have described a new reality within optoelectronics and spintronics [1-10]. The intrinsic and extrinsic characteristics [11, 12] of some magnetic materials under the influence of light can produce unimaginable effects [13-15]. The main magneto-optical interactions were discovered by Faraday, which gave rise to the Faraday effect [16, 17], and by Kerr, which gave rise to the Kerr effect [18, 19]. Static behavior has always been dominant in this type of system, as to date, no dynamic properties have been reported regarding the mutual dependence of light [20, 21] and magnetic matter [22, 23, 24]. Light represents a disturbance external to materials, which causes electromagnetic modes to be obscured in ferromagnetic materials. This description works well in most cases because the light-magnetic matter interaction is assumed to be weak. With increasing light-magnetic matter interaction, however, it has reached a regime in which electromagnetic states fundamentally alter the magneto-optical interactions of materials [25, 26, 27, 28]. Such interactions are usually quantified using the magneto-optical states: when the predominant magneto-optical interactions are demagnetizing (PMOID), or when the predominant magneto-optical interactions are magnetizing (PMOIM) [1, 2, 6, 8].

\setlength{\parindent}{25pt}The magnetization of nanostructures plays a key role when it is interested in seeking answers regarding magneto-optical interactions, where the result will depend on the extrinsic and intrinsic properties of the materials used [11, 12]. Nanostructures have great applicability and have been used as a means to manipulate these interactions for the creation of new devices. Some studies have been improving already known technologies, such as optical sensors, and enabling the creation of new devices with applications in biomedicine, for example [29, 30]. When addressing issues involving magnetic interactions, we must emphasize the importance of the system characterized as nanostructured, requiring the observation of components considered reversible and irreversible due to the application of the magnetic field (H).

\setlength{\parindent}{25pt}The effects of magnetic interactions are important for quantum devices since they can influence the spin properties of the nanostructures, as nanowire arrays. The synthesis of nanowires through electrochemistry [31] has already been studied, but nothing has been measured regarding the interactions that occur. In applications, electrodeposition on aluminum membranes is the most used for the development of systems and devices [5, 6, 8, 9]. This type of system can present different magnetization reverse modes [32, 33, 34]. Such systems can be strongly influenced by magneto-optical interactions [1, 2]. The influence is detected mainly during the magnetization process with light, which always presents reversible and irreversible components. Furthermore, a striking feature of the magnetization process is that it is not possible to separate the parts of their hysteresis without losing information due to changes in the magnetic energies of the structure. This means that many properties remain hidden during the magnetization process, and there is a need for understanding. The study from the magneto-optical interactions can be performed using the remanent state obtained during the magnetization process [1, 2].

In a particular system, the well-established normalized $\Delta m$ curves ($\Delta m_{N}$) produce results of the interaction effects; such curves are comparisons between optical isothermal remanent magnetization (OIRM(H)) and optical direct current demagnetization (ODCD(H)) curves, which defines other physical quantities such as m$_{d}$(H) = ODCD(H)/OIRM(H$_{Max})$ and m$_{r }$(H) = OIRM(H)/OIRM(H$_{Max})$ that are normalized considering the value obtained with maximum magnetic field [1-8, 35-38]. In this paper, we present an experimental and simulated study on the predominant magneto-optical interactions in nanostructures, for that, we perform experiments and simulations in a system of magnetic nanowire arrays in alumina membranes. After analyzing magneto-optical interactions, we observe two types of magnetic states, i.e., magnetized and demagnetized, which reveal the main characteristics of magneto-optical energies. Our studies seek to describe the light-matter interaction with the application of a magnetic field, where the modulated white light only serves to excite the magneto-optical effects, that is, all results are obtained considering the dependence of magneto-optical interactions with the magnetic field.

\section{Experimental section}

\subsection{Sample manufacturing}

For the preparation of hexagonal arrays of magnetic nanowires were used membranes of anodic aluminum oxide (AAO) as templates and electrodeposited the desired materials (cobalt-Co, iron-Fe, and nickel-Ni) within the cylindrical pores. AAO was obtained by electrochemical oxidation of aluminum plates Aldrich 99.9999$\%$, with a voltage of 20 V in aqueous acid solutions. The experiments were carried out using a Potentiostat Model IVIUMSTAT.XRe. As a result, we obtained cylindrical pores diameter of 27 nm, and a center-to-center pores distance of 55 nm, as shown in \textbf{Fig. 1}. The nanowires were manufactured by electrodeposition with AC potential of 17 V$_{rms}$, which have an average length of 2 $\mu$m.

\subsection{Experimental setup of magneto-optical measurements}

We performed magnetic measurements using the Vibrating Sample Magnetometry (VSM) method, one of the systems used to study magnetic properties [35-38]. We used a VSM model MicroSense EV7 with a magnetic detection sensitivity in order from 10$^{-7}$ emu. The optical properties of the samples have been studied with a magneto-optical absorption spectrometer with the schematics shown in \textbf{Fig. 1}. The collimated white light from a 1000 W Hg(Xe) DC arc lamp is modulated by a chopper with a frequency of 1 MHz, and amplitude of 10 W and focused on the sample, where magneto-optical interactions are measured via VSM. Part of the reflected signal was captured by a photodetector, which was used only to observe whether there was any voltage loss during the process. As discussed, $\Delta m_{N}$ give results of the interaction effects, such curves are comparisons between optical isothermal remanent magnetization (OIRM(H)) and optical direct current demagnetization (ODCD(H)) curves [1-8, 35-38]. In an OIRM(H) measurement, the starting point is the demagnetized sample after cooling in a zero magnetic field. Then, a small magnetic field is applied after a time interval sufficient for the magnetic equilibrium. Next, the small magnetic field is switched off and the remanence is measured. The process is repeated, increasing the value of the magnetic field until the sample reaches saturation and remanence takes its higher value. The ODCD(H) measurement is similar to the OIRM(H). However, the sample is initially in a saturated state. Then, the applied magnetic field is slowly inverted until a small value opposite to the initial magnetization is identified. After the applied magnetic field is switched off and the remanence measured. The process is repeated step by step, increasing the value of the inverted magnetic field until saturation. These processes determine the $\Delta m_{N}$ in magnetic samples [35-39] and are in accord with Henkel's proposal [35-37] and J. Holanda's models [1, 2]. On the other hand, if magnetic interactions are not present, the magnetic system is described by the Stoner and Wohlfarth proposal [40].

\begin{figure}[h]
	\vspace{0.5mm} \hspace{0.5mm}
	\begin{center}
		\includegraphics[scale=0.23]{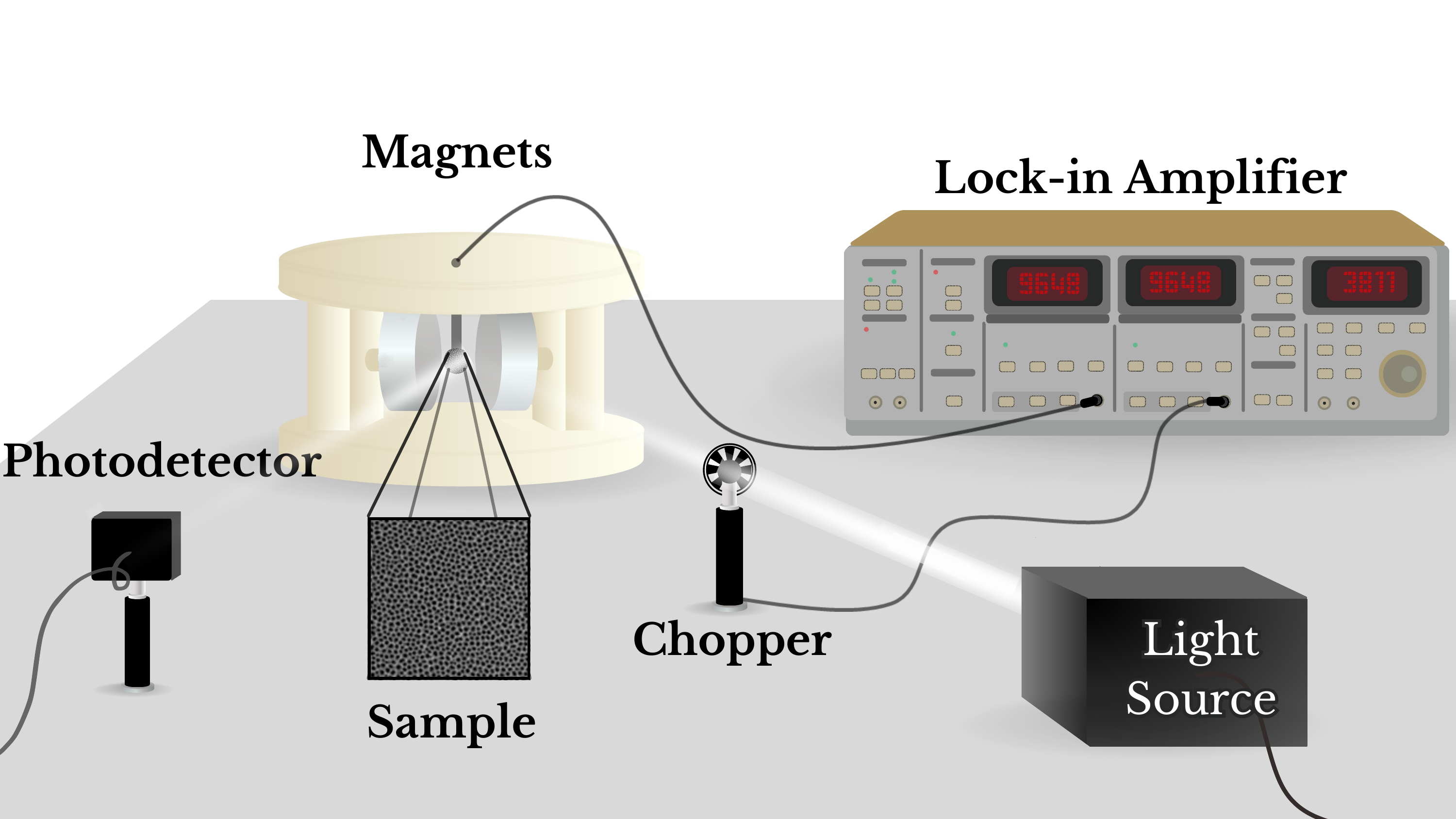}
		\caption{\label{arttype}(Color on-line) Experimental setup used to measure the optical isothermal remanent magnetization (OIRM(H)) and optical direct current demagnetization (ODCD(H)) curves, and consequently obtain the $\Delta m_{N}$ curves as a function of the magnetic field (H).}
		\label{tatu}
	\end{center}
\end{figure}

\section{Experimental and simulated results}

\subsection{Sample – Cobalt (Co)}

We started our studies on cobalt (Co) nanowire arrays, as shown in \textbf{Fig. 2(a)}. We performed the $\Delta m_N$ measurements without (blue) and with (red) the incidence of modulated light, which presented intensities for the PMOID and PMOIM states. The PMOID state presented the following intensities, $I_{M}^{D-Ex} (a. u.)_{Co}$ = -1.40 and $I_{M-MO}^{D-Ex} (a. u.)_{Co }$ = -1.70 for the purely demagnetizing and magneto-optical demagnetizing, respectively; this allowed obtaining a purely optical demagnetizing intensity $I_{O}^{D-Ex} (a. u.)_{Co}$ = -0.34. Similarly, the PMOIM state presented the following intensities, $I_{M}^{M-Ex} (a. u.)_{Co}$ = 0.16, and $I_{M-MO}^{M-Ex} (a. u.)_ {Co}$ = 0.11 for the purely magnetizing and magneto-optical magnetizing, respectively; this allowed obtaining a purely optical magnetizing intensity $I_{O}^{M-Ex} (a. u.)_{Co}$ = -0.05. To unequivocally prove the effects resulting from the light-magnetic matter interaction, we carried out simulations using the models presented in references [1] and [2] and confirmed the results obtained experimentally, as presented in \textbf{Fig. 2(b)} and \textbf{Table 1}.

\begin{figure}[h]
	\vspace{0.5mm} \hspace{0.5mm}
	\begin{center}
		\includegraphics[scale=0.68]{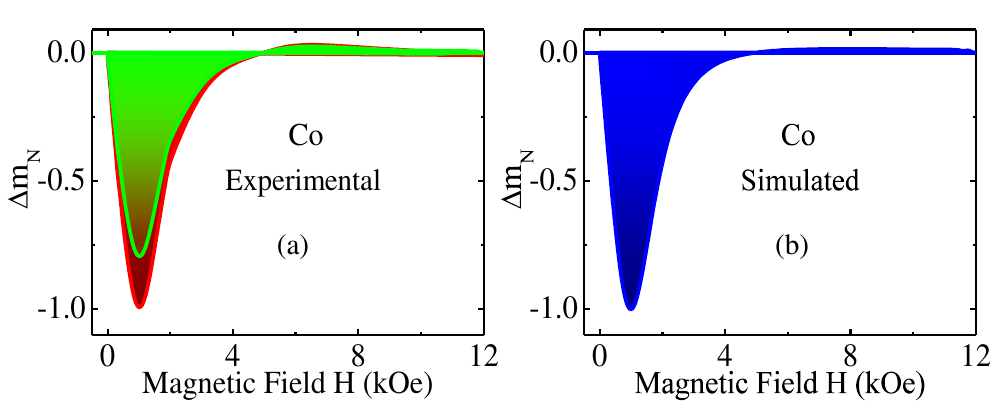}
		\caption{\label{arttype}(Color on-line) Experimental $\Delta m_{N}$ curves of cobalt (Ni) nanowire arrays as a function of magnetic field (H), which were obtained considering the optical isothermal remanent magnetization (OIRM(H)) and optical direct current demagnetization (ODCD(H)) curves. (a) Experimental. (b) Theoretical.}
		\label{patu}
	\end{center}
\end{figure}

Optical interactions arise from the coupling of the electromagnetic field of light with the domain walls existing between the monodomain grains in nanowires [1-8]. The phenomenon is sensitive to the static magnetic field and proportional to the remanent magnetization of the material and the intensity of the incident light. In cobalt nanowires, where the predominant magnetization reversal modes are curling and buckling [3-11, 14-17, 34-40], the effects from the light-magnetic matter interaction are due to the curling and buckling domain walls. The curling mode occurs around the nanowire axis, allowing a three-dimensional closed flow, where the spins progressively invert through the propagation of a vortex. The buckling mode is similar to the coherent one; the magnetization is also uniform for any plane perpendicular to the nanowire axis. However, there is additionally a periodicity from the magnetization distribution along the nanowire axis. The behavior from the phenomenon allows an energetic variation in the dipolar and exchange regime, which highlights the new magneto-optical behavior.

\begin{table}[h]
	\caption{\label{tab:table4}{Shows the intensity values of interactions obtained for the Co samples, where the values from the intensities of the purely magnetic (PM), magneto-optical (MO) and optical (O) interactions are presented for the two magnetic states, i.e., the magnetizing and demagnetizing states.}}%
	\begin{center}
		\begin{tabular}{ccc|c}
			& Experimental &  & Simulated \\
			\hline
			\hline
			PM interactions & 	MO interactions & O interactions & MO interactions \\
			\hline
			$I_{M}^{D-Ex} (a. u.)_{Co}$ & $I_{M-MO}^{D-Ex} (a. u.)_{Co}$ & $I_{O}^{D-Ex} (a. u.)_{Co}$ & $I_{M-MO}^{D-Si} (a. u.)_{Co}$ \\
			\hline
			-1.40 & -1.74 & -0.34 & -1.70 \\
			\hline
			$I_{M}^{M-Ex} (a. u.)_{Co}$ & $I_{M-MO}^{M-Ex} (a. u.)_{Co}$ & $I_{O}^{M-Ex} (a. u.)_{Co}$ & $I_{M-MO}^{M-Si} (a. u.)_{Co}$ \\
			\hline
			0.16 & 0.11 & -0.05 & 0.08 \\
		\end{tabular} 
	\end{center}
\end{table}

\subsection{Sample – Iron (Fe)}

We continued our studies on iron (Fe) nanowire arrays, as shown in Fig. 2(a). We carried out measurements of $\Delta m_N$ without (blue) and with (red) the incidence of modulated light, which presented intensities for the PMOID and PMOIM states in the same way as the Co samples. The PMOID state presented the following intensities, $I_{M}^{D-Ex} (a. u.)_{Fe}$ = -1.05, and $I_{M-MO}^{D-Ex} (a. u.)_{ Fe}$ = -1.18 for the purely demagnetizing and magneto-optical demagnetizing, respectively; this allowed obtaining a purely optical demagnetizing intensity $I_{O}^{D-Ex} (a. u.)_{Fe}$ = -0.13. Concomitantly, the PMOIM state presented the following intensities, $I_{M}^{M-Ex} (a. u.)_{Fe}$ = -1.05, and $I_{M-MO}^{M-Ex} (a. u.) _{Fe}$ = -1.18 for the purely magnetizing and magneto-optical magnetizing, respectively; which also allowed us to obtain a purely optical magnetizing intensity $I_{O}^{M-Ex} (a. u.)_{Fe}$ = -0.13. To unequivocally prove the effects resulting from the light-magnetic matter interaction, we also carried out simulations using the models presented in references [1] and [2] and confirmed the results obtained experimentally, as presented in \textbf{Fig. 3(b)} and \textbf{Table 2}. 

\begin{figure}[h]
	\vspace{0.5mm} \hspace{0.5mm}
	\begin{center}
		\includegraphics[scale=0.68]{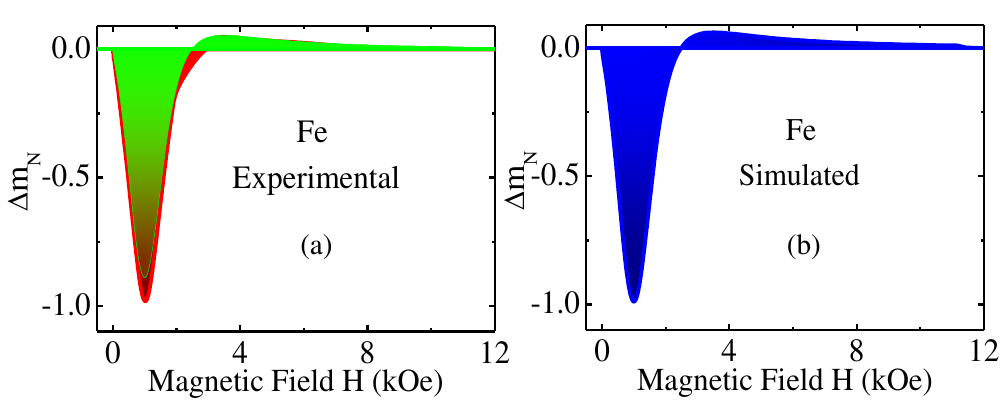}
		\caption{\label{arttype}(Color on-line) Experimental $\Delta m_{N}$ curves of iron (Fe) nanowire arrays as a function of magnetic field (H), which were obtained considering the optical isothermal remanent magnetization (OIRM(H)) and optical direct current demagnetization (ODCD(H)) curves. (a) Experimental. (b) Theoretical.}
		\label{patu}
	\end{center}
\end{figure}

Optical interactions also arise from the coupling of the electromagnetic field of light with the domain walls existing between the monodomain grains in the nanowires [3-11, 14-17, 33-40]. The phenomenon is sensitive to the static magnetic field and proportional to the remanent magnetization of the material and the intensity of the incident light. In iron nanowires, where the predominant magnetization reversal modes are transverse and vortex [3-11, 14-17, 33-40], the effects from the light-magnetic matter interaction are due to the transverse and vortex domain walls. In the transverse mode, the spins are progressively inverted through the propagation of a transverse domain wall, while the vortex mode represents the localized curling mode. Similar to the results obtained with Co samples, we found a behavior of the phenomenon that allows an energetic variation in the dipolar and exchange regime, showing the importance of magneto-optical effects.

\begin{table}[h]
	\caption{\label{tab:table4}{Shows the intensity values of interactions obtained for the Fe samples, where the values of the intensities of the purely magnetic (PM), magneto-optical (MO), and optical (O) interactions are presented for the two magnetic states, that is, the magnetizing and demagnetizing states.}}%
	\begin{center}
		\begin{tabular}{ccc|c}
			& Experimental &  & Simulated \\
			\hline
			\hline
			PM interactions & 	MO interactions & O interactions & MO interactions \\
			\hline
			$I_{M}^{D-Ex} (a. u.)_{Fe}$ & $I_{M-MO}^{D-Ex} (a. u.)_{Fe}$ & $I_{O}^{D-Ex} (a. u.)_{Fe}$ & $I_{M-MO}^{D-Si} (a. u.)_{Fe}$ \\
			\hline
			-1.05 & -1.18 & -0.13 & -1.16 \\
			\hline
			$I_{M}^{M-Ex} (a. u.)_{Fe}$ & $I_{M-MO}^{M-Ex} (a. u.)_{Fe}$ & $I_{O}^{M-Ex} (a. u.)_{Fe}$ & $I_{M-MO}^{M-Si} (a. u.)_{Fe}$ \\
			\hline
			0.23 & 0.13 & -0.10 & 0.26 \\
		\end{tabular} 
	\end{center}
\end{table}

\subsection{Sample – Nickel (Ni)}

We finalize our studies considering nickel (Ni) nanowire arrays, as shown in \textbf{Fig. 4(a)}. We performed the $\Delta m_N$ measurements without (blue) and with (red) the incidence of modulated light, which presented intensities for the PMOID and PMOIM states. The PMOID state presented the following intensities, $I_{M}^{D-Ex} (a. u.)_{Ni}$ = -1.78, and $I_{M-MO}^{D-Ex} (a. u.)_{ Ni}$ = -1.98 for the purely demagnetizing and magneto-optical demagnetizing, respectively; this allowed obtaining a purely optical demagnetizing intensity $I_{O}^{D-Ex} (a. u.)_{Ni}$ = -0.20. Similarly, the PMOIM state presented the following intensities, $I_{M}^{M-Ex} (a. u.)_{Ni}$ = 0.14, and $I_{M-MO}^{M-Ex} (a. u.)_ {Ni}$ = 0.12 for the purely magnetizing and magneto-optical magnetizing, respectively; this allowed obtaining a purely optical magnetizing intensity $I_{O}^{M-Ex} (a. u.)_{Ni}$ = 0.02. In the same way, as for the other materials (Co and Fe), we carried out simulations using the models presented in references [1] and [2] and confirmed the results obtained experimentally, as presented in \textbf{Fig. 4(b)} and \textbf{Table 3}.

\begin{figure}[h]
	\vspace{0.5mm} \hspace{0.5mm}
	\begin{center}
		\includegraphics[scale=0.68]{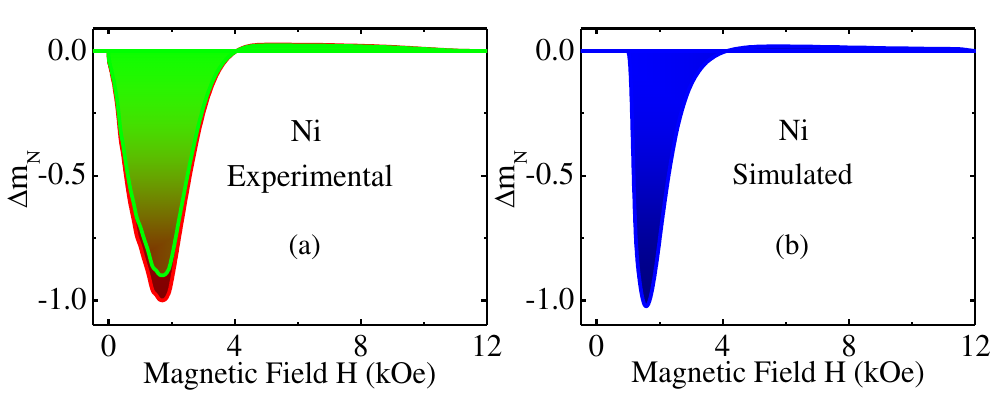}
		\caption{\label{arttype}(Color on-line) Experimental $\Delta m_{N}$ curves of iron (Fe) nanowire arrays as a function of magnetic field (H), which were obtained considering the optical isothermal remanent magnetization (OIRM(H)) and optical direct current demagnetization (ODCD( H)) curves. (a) Experimental. (b) Theoretical.}
		\label{patu}
	\end{center}
\end{figure}

As discussed for the previous materials (Co and Fe), optical interactions in Ni also arise from the coupling of the electromagnetic field of light with the domain walls existing between the monodomain grains in the nanowires [1-3, 10, 12, 29, 34]. In Ni nanowires, the phenomenon is also sensitive to the static magnetic field and proportional to the remanent magnetization of the material and the intensity of the incident light. In this type of nanowires, where the predominant mode of magnetization reversal is coherent [1-3, 10, 12, 29, 34], the effects from the light-magnetic matter interaction are due to the domain walls between the coherent modes of the grains. In this case, the grains that are pure monodomains are misaligned with the others, which means that the interaction effects are purely due to the reversal between grains. We emphasize that the coherent mode represents unison rotation for all spins. Furthermore, we emphasize that the behavior from the phenomenon allowed an energetic variation in the dipolar regime and exchange between grains, which shows a magneto-optical behavior of interface. In this case, our experimental and simulated results had a considerable discrepancy because, in the model, we considered an approach similar to the previous ones, where the energetic variation also has a contribution from within the grains.

\begin{table}[h]
	\caption{\label{tab:table4}{Shows the intensity values of interactions obtained for the Ni samples, where the values from the intensities of the purely magnetic (PM), magneto-optical (MO), and optical (O) interactions are presented for the two magnetic states, that is, the magnetizing and demagnetizing states.}}%
	\begin{center}
		\begin{tabular}{ccc|c}
			& Experimental &  & Simulated \\
			\hline
			\hline
			PM interactions & 	MO interactions & O interactions & MO interactions \\
			\hline
			$I_{M}^{D-Ex} (a. u.)_{Ni}$ & $I_{M-MO}^{D-Ex} (a. u.)_{Ni}$ & $I_{O}^{D-Ex} (a. u.)_{Ni}$ & $I_{M-MO}^{D-Si} (a. u.)_{Ni}$ \\
			\hline
			-1.78 & -1.98 & -0.20 & -1.21 \\
			\hline
			$I_{M}^{M-Ex} (a. u.)_{Ni}$ & $I_{M-MO}^{M-Ex} (a. u.)_{Ni}$ & $I_{O}^{M-Ex} (a. u.)_{Ni}$ & $I_{M-MO}^{M-Si} (a. u.)_{Ni}$ \\
			\hline
			0.12 & 0.14 & 0.02 & 0.10 \\
		\end{tabular} 
	\end{center}
\end{table}

The behaviors observed for the three materials (cobalt-Co, iron-Fe, and nickel-Ni) reveal a high energy balance on both the behaviors: dipolar and exchange. This fact characterizes observation and opens up the most diverse possible applications for our experimental and theoretical description. Furthermore, the fact that the samples have a high degree of density of magnetic material distributed evenly in small regions opens up the possibility of this type of system being very important for so-called intelligent energy systems. It is also possible to assume that such dynamic observations can generate a spin-optoelectronic current with characteristics similar to a spin-polarized current with four degrees of freedom, two due to spin effects and the other two due to optical effects.

\section{Conclusion}

The experimental behavior of magneto-optical interactions in our magnetic structures revealed two predominant magneto-optical states: PMOID and PMOIM. Understanding how each state arises due to the different effects produced during the magnetization process makes its study of fundamental importance for device applications in areas such as quantum computing and engineering. Our results also consolidate an efficient way to describe the behavior of magneto-optical interactions, directly considering that from interactions with the magnetic field during the light-magnetic matter interaction process. Furthermore, the results show that the effects that cause global magnetic states (demagnetized and magnetized) influence the excitation of magneto-optical waves using the conventional method or light-matter interaction. In terms of fundamentals, we demonstrate a significant advance in understanding the behavior of fundamental interactions in magnetic structures under the influence of light.

\section*{Author contributions}

T. Ferro, L. Hildever, A. Dias, A. José, and F. Estrada performed and analyzed all the experimental measures. And J. Holanda supervised, wrote and discussed the work with everyone.

\section*{Conflicts of interest}

The authors declare that they have no conflict of interest.

\section*{Data availability statement}

The data generated and/or analysed during the current study are not publicly available 
for legal/ethical reasons but are available from the corresponding author on reasonable 
request.

\section*{Acknowledgements} 

This research was supported by Conselho Nacional de Desenvolvimento Científico e Tecnológico (CNPq), Coordenação de Aperfeiçoamento de Pessoal de Nível Superior (CAPES), and Fundação de Amparo à Ciência e Tecnologia do Estado de Pernambuco (FACEPE). The authors would like to thank researcher Chenbo Zhao from Lanzhou University for his help in preparing the samples, and to the Centro Multiusuário de Pesquisa e Caracterização de Materiais da Universidade Federal Rural de Pernambuco (CEMUPEC-UFRPE). 

\bibliographystyle{MiKTeX}

\end{document}